%% file: main.tex
\documentclass[sigconf, nonacm]{acmart}
\usepackage{amsmath,amsfonts}
\usepackage{algorithm}
\usepackage{algorithmic}
\usepackage{booktabs}
\usepackage{graphicx}
\usepackage{textcomp}

\usepackage{xcolor}
\usepackage{url}
\usepackage{stfloats} 
\usepackage{soul}

\usepackage{multirow}
\usepackage{graphics}
\usepackage{enumitem}
\usepackage{array}
\usepackage{makecell}
\usepackage{amsthm}
\usepackage{subcaption}

\AtBeginDocument{%
  \providecommand\BibTeX{{%
    Bib\TeX}}}

\copyrightyear{2026}
\acmYear{2026}
\setcctype{by}
\acmConference[WWW '26] {Proceedings of the ACM Web Conference 2026}{April 13--17, 2026}{Dubai, United Arab Emirates.}
\acmBooktitle{Proceedings of the ACM Web Conference 2026 (WWW '26), April 13--17, 2026, Dubai, United Arab Emirates}
\acmISBN{979-8-4007-2307-0/2026/04}
\acmDOI{10.1145/3774904.3792167}

\begin{document}

\author{Xianquan Wang}
\authornote{Equal Contribution.}
\affiliation{
  \institution{University of Science and Technology of China}
  \city{Hefei}
  \country{China}
}
\email{wxqcn@mail.ustc.edu.cn}

\author{Zhaocheng Du}
\authornotemark[1]
\affiliation{
  \institution{Huawei Noah's Ark Lab}
  \city{Shenzhen}
  \country{China}
}
\email{zhaochengdu@huawei.com}

\author{Jieming Zhu}
\affiliation{
  \institution{Huawei Noah's Ark Lab}
  \city{Shenzhen}
  \country{China}
}
\email{jamie.zhu@huawei.com}

\author{Qinglin Jia}
\affiliation{
  \institution{Huawei Noah's Ark Lab}
  \city{Beijing}
  \country{China}
}
\email{jiaqinglin2@huawei.com}

\author{Zhenhua Dong}
\affiliation{
  \institution{Huawei Noah's Ark Lab}
  \city{Shenzhen}
  \country{China}
}
\email{dongzhenhua@huawei.com}

\author{Kai Zhang}
\authornote{Corresponding Author.}
\affiliation{
  \institution{University of Science and Technology of China}
  \city{Hefei}
  \country{China}
}
\email{kkzhang08@ustc.edu.cn}

\renewcommand{\shortauthors}{Xianquan Wang et al.}

\def\BibTeX{{\rm B\kern-.05em{\sc i\kern-.025em b}\kern-.08em
    T\kern-.1667em\lower.7ex\hbox{E}\kern-.125emX}}

\title{FairFS: Addressing Deep Feature Selection Biases for Recommender System}

\begin{abstract}
Large-scale online marketplaces and recommender systems serve as critical technological support for e-commerce development. In industrial recommender systems, features play vital roles as they carry information for downstream models. Accurate feature importance estimation is critical as it helps find the most useful feature subsets from thousands of feature candidates for online services. With such a selection, optimizing online performance while reducing computation burden is possible. To solve the feature selection problems of deep learning, trainable gate-based and sensitivity-based methods are proposed and proven effective in the industry. Nevertheless, by analyzing real-world examples, we identified three bias issues that make feature importance estimation rely on partial model layers, samples, or gradients to make decisions, ultimately leading to an inaccurate feature importance estimation. We call these biases layer bias, baseline bias, and approximation bias. To mitigate these three biases, we propose \textbf{FairFS}, a \textbf{fair} and \textbf{accurate} feature selection algorithm. On one hand, FairFS directly regularizes feature importance estimated across all non-linear transformational layers to avoid layer bias. On the other hand, it utilizes a smooth baseline feature that is close to the classifier's decision boundary and an aggregated approximation method to mitigate bias issues. Extensive experiments show how FairFS mitigates these three biases and achieves SOTA feature selection results.
\end{abstract}


\maketitle

\input{sections/introduction}

\input{sections/relatedwork}
\input{sections/method.tex}
\input{sections/experiments.tex}
\input{sections/conclusion.tex}

\section*{Acknowledgement}
This research was partially supported by the National Natural Science Foundation of China (Grants No.62406303, No.U25B2072) and Anhui Province Science and Technology Innovation Project (No.202423k09020010).
 
\bibliographystyle{ACM-Reference-Format}
\bibliography{references}

\input{sections/appendix.tex}

\end{document}

%% file: sections/introduction.tex
\section{Introduction}

Deep neural networks~\cite{dnn, li2024cetn} have become the fundamental backbones of the entire pipeline for recommender systems~\cite{cold, youtubenet,rerank,xdeepfm, li2025revisiting,li2024simcen}.
To provide the input information for the neural network, engineers have to manually construct features that can represent users' and items' characteristics~\cite{featureengineering,zhang2025simcdr,zhang2021multi,sweet}. 
In industry, thousands of features~\cite{openfe} might be built in order to improve their comprehensiveness.
Nevertheless, due to two practical issues, these feature columns (\textit{a.k.a.} feature field, explained in Sec.~\ref{note}) cannot be haphazardly packed into neural networks for online service. 
One is that the noisy information present in these features will waste the expressive power of neural networks and deteriorate model performance~\cite{dealing,liu2025elo,liu2024computerized}.
\begin{figure}[!t]
  \centering
    \includegraphics[width=0.85\linewidth]{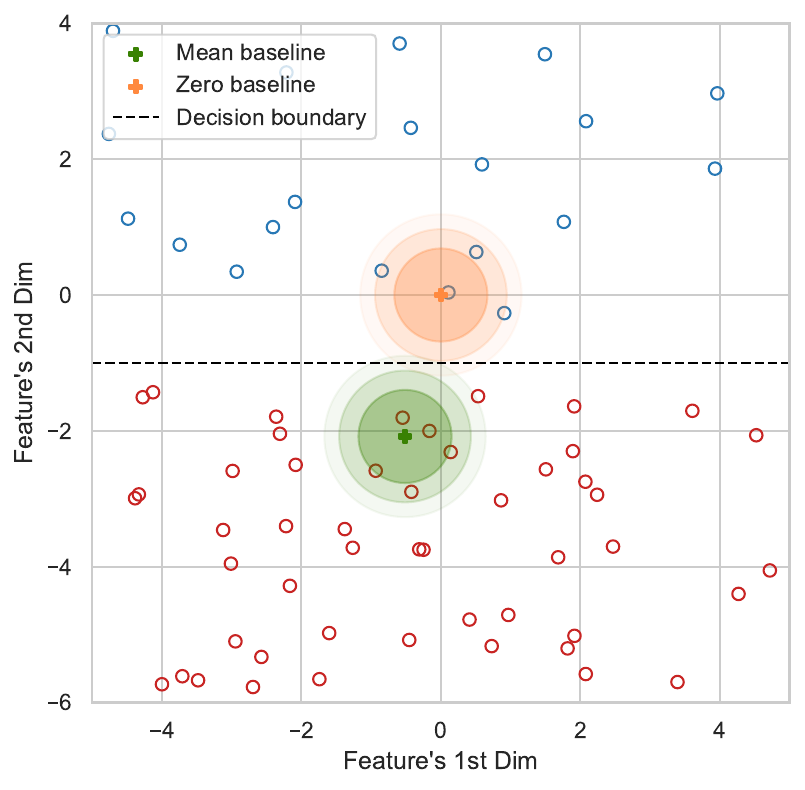}
  \caption{Baseline Bias: In imbalanced binary classification, both mean and zero baselines are biased—zero skews toward positive samples, and mean toward negative ones—causing biased feature importance estimates.}
  \label{data_bias}
  \vspace*{-0.5cm}
\end{figure}
The other is that redundant features lead to higher serving latency~\cite{pessimistic,li2025madakv,li2025mergenet}. 
Consequently, feature selection turns into a necessary procedure for industrial recommendations.

To perform \textbf{feature field} selection, it is essential to estimate field importance~\cite{du2024tutorial,du2024lightcs,jia2024erase,wang2025tayfcs,wang2025mitigating}, which measures each feature's contribution to model accuracy~\cite{jia2025self}. Traditional methods can be broadly categorized as filter-based  (\textit{e.g.}, Person Correlation~\cite{nasir2020pearson}, MRMR~\cite{radovic2017minimum}), wrapper-based (RFE~\cite{chen2007enhanced}), and embedded-based (\textit{e.g.}, Lasso~\cite{fonti2017feature}, Xgboost~\cite{chen2020improving}) approaches. While filter-based methods rely on statistical measures and wrapper-based methods involve recursive optimization, embedded-based methods integrate feature selection directly into model training, making them more suitable for capturing complex non-linear relationships in deep learning tasks. In recent years, embedded-based methods have become the dominant approach in deep feature selection research. Detailed classifications and methodologies are discussed in the Related Work section. Based on definitions of feature importance, embedded-based methods can be further divided into the following two types.

The first is Trainable Gate-based Methods (\cite{autofield,lassonet,stg,du2024lightcs}): These methods define feature importance using the absolute values of regularized feature gate parameters, where larger gate values amplify feature signals, highlighting their role in predictions. However, non-linear transformations in subsequent layers can distort this correlation, favoring features with larger initial gate values while neglecting downstream influences. We call it \textbf{\textcircled{1}Layer Bias}.

The second is Sensitivity-based Methods (\cite{shark, multisfs, permutation, lofo}): These methods quantify feature importance by measuring the difference in model loss when a feature is informative versus non-informative (baseline feature). The magnitude of this loss difference reflects the feature's contribution to optimization. However, defining the baseline feature remains debated—some use zero embeddings~\cite{snip}, while others prefer mean embeddings~\cite{mask}. Poor baseline choices can lead to \textbf{\textcircled{2}Baseline Bias}, where samples near the baseline are misclassified as non-informative, skewing feature importance estimates (see Figure~\ref{data_bias}).

Furthermore, in consideration of efficiency in industrial scenarios, modern sensitivity-based methods approximate loss sensitivity with $1^{st}$ order Taylor approximation. This approximation relies on the assumption of linearity in the loss curve between the baseline feature and the target feature, an assumption that is inherently incorrect. We call this bias as \textbf{\textcircled{3}Approximation Bias}.

These three biases—layer bias, baseline bias, and approximation bias—affect the accuracy of feature importance estimation. To address them, we introduce a simple, model-agnostic framework called \textbf{FairFS} (unbiased Feature Importance Regularization). To alleviate the \textcircled{1}\ul{layer bias}, we directly sparsify sensitivity-based feature importance as a loss term, leveraging the loss sensitivity to account for all non-linearities. For the \textcircled{2}\ul{baseline bias}, we establish two principles that non-informative baseline features should meet, from which we derive a closed-form solution and a simplified substitute. Regarding the \textcircled{3}\ul{approximation bias}, we adopt an aggregated approximation strategy, dividing the original approximation into multiple equidistant points, calculating gradient for  each one individually, and then combining the results to achieve a more accurate estimation of loss sensitivity. As the number of points approaches infinity, the approximation bias approaches zero. These strategies mitigate the biases inherent in current feature importance estimation algorithms, and FairFS demonstrates state-of-the-art performance in feature selection for recommender systems.

To summarize, our main contributions can be listed as follows:
\begin{itemize}[leftmargin=*]
    \item We identify and illustrate three kinds of biases (\textcircled{1}\textcircled{2}\textcircled{3}) in current feature importance estimation methods with examples.
    \item We design a simple and model agnostic framework to mitigate three bias issues mentioned above with theoretical support.
    \item We have formulated two mathematical principles that a baseline feature should adhere to. 
    \item Extensive experiments on public datasets and one A/B test on industry ads platform have verified the superiority of our method.
\end{itemize}

The code is available at~\url{https://github.com/xqwustc/FairFS/} for ease of reproduction. 

\begin{figure}[!t]
  \centering
    \includegraphics[width=1\linewidth]{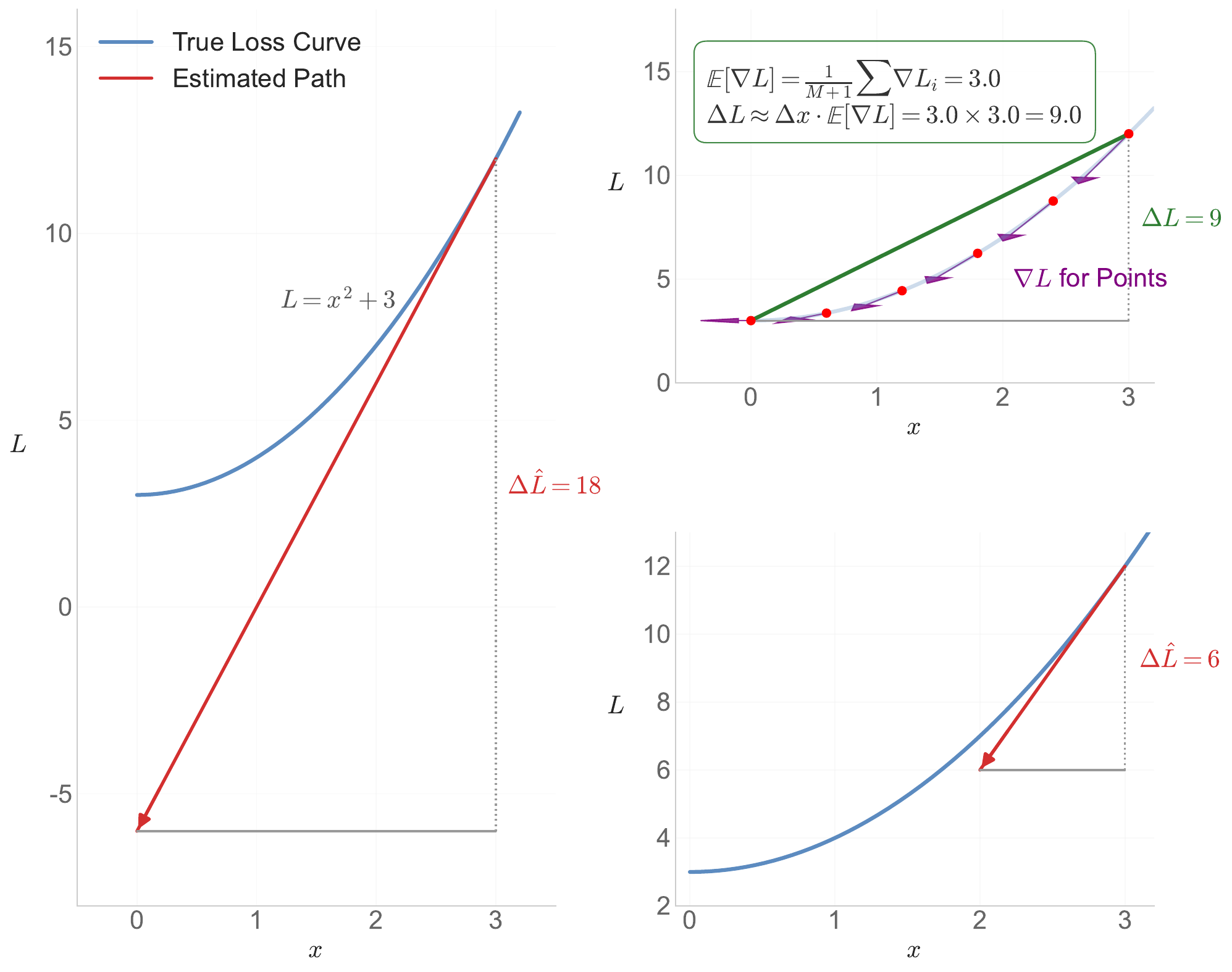}
\vspace*{-0.35cm}
  \caption{Approximation bias: We examine feature $x$'s importance to loss $L=x^2+3$ by measure loss change when switching $x$ from informative to non-informative (zero). SHARK (left) and SFS (right bottom) have larger approximation error than our aggregated approximation method (right top).}
  \label{Approximation_bias}
\end{figure}

%% file: sections/relatedwork.tex
\section{Related Work}
\label{Related}
\subsection{Feature selection for shallow models}
Feature selection methods based on interaction paradigms with ML models are categorized into filter-based, wrapper-based, and embedded-based approaches: Filter-based methods determine feature subsets using statistical measures like Pearson Correlation~\cite{nasir2020pearson}, MRMR~\cite{radovic2017minimum}, or VIF~\cite{cheng2022variable}, without interacting with ML models. They are unsuitable for deep learning tasks due to their inability to capture non-linear relationships. Wrapper-based methods iteratively include or eliminate features using ML models, as seen in RFE~\cite{chen2007enhanced}. However, their computational complexity limits their scalability to large datasets. Embedded-based methods, such as Lasso~\cite{fonti2017feature} and XGBoost~\cite{chen2020improving}, integrate feature importance estimation into model training. These approaches are well-suited to neural networks, addressing non-linearity despite computational costs. Notably, about 70\% of deep feature selection algorithms proposed in the past five years fall into this category. The following subsections detail two subtypes of embedded-based deep feature selection methods.

\subsection{Trainable Gate-Based Method}
These methods assign a trainable gate to each feature field, using the absolute value of the trained gate parameter to determine feature importance—larger values indicate greater importance. Early approaches like LassoNet~\cite{lassonet} regularized the first weight matrix following the feature and eliminated sparse weights. Later methods, such as Droprank~\cite{droprank}, STG~\cite{stg}, LPFS~\cite{lpfs}, AutoField~\cite{autofield}, and FSCD~\cite{fscd}, replaced weight matrices with individual gate parameters and employed techniques like Gumbel Softmax~\cite{gumbel} or Proximal Gradient Descent~\cite{proximal} for discrete feature selection. Soft-gate methods, including AdaFS~\cite{adafs} and MvFS~\cite{mvfs}, avoided hard discretization by using weight multiplications during inference.

\begin{figure}[!t]
  \centering
    \includegraphics[width=1\linewidth]{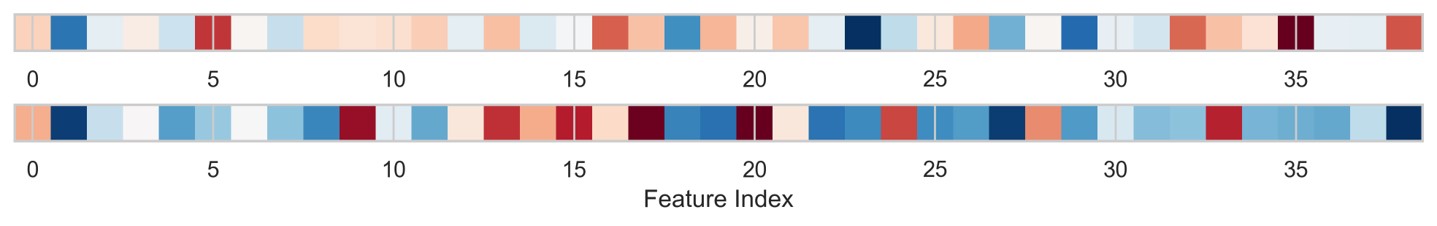}
  \vspace*{-0.4cm}
  \caption{Layer bias: The top shows feature gate magnitudes, and the bottom shows their changes after a hidden layer (square matrix to avoid weight mixing). Colors range from red (low) to blue (high).} 
  \label{Layer_bias}
  \vspace{-1.5em}
\end{figure}

\textbf{$\cdot$ Layer Bias}. Despite their simplicity, these methods lack theoretical robustness, particularly in multi-layer perceptrons (MLPs), where non-linear transformations can diminish the influence of gate values. As a result, they often favor features with larger initial gate parameters. To investigate, we analyzed a pre-trained Droprank model on the Criteo dataset~\cite{autoint}, focusing on the first transformation layer, where features are treated individually before subsequent non-linear combination. Figure \ref{Layer_bias} reveals a mismatch between the gating layer's amplification scale and that of the subsequent transformation, confirming the presence of layer bias.

\subsection{Sensitivity-Based Method}
These methods define feature importance as the difference in loss when a target feature is replaced with a baseline feature. While direct computation through methods like Leave-One-Out FS~\cite{lofo} and Permutation Feature Importance (PFI)~\cite{permutation} is accurate, it is computationally expensive, as each feature requires an additional evaluation step. To improve efficiency, modern approaches rely on gradient-based approximations. SFS~\cite{multisfs} estimates feature importance using gradients, while SHARK~\cite{shark} refines this with first-order Taylor expansion, multiplying the gradient by the distance between the feature and a mean baseline. Similarly, SNIP~\cite{snip} applies the same method with a zero baseline. Despite their efficiency, these methods face two key biases:

\textbf{ $\cdot$ Baseline Bias}. Various baselines, such as zero, mean, or permuted values, are used to represent non-informative features. However, they often fail to align with the principles of non-informativeness in recommendation systems. For instance, as shown in Section~\ref{Baseline_Analysis}, \textbf{the mean baseline in the Criteo dataset produces logits of 0.293, far from the 0.5 decision boundary}. This misclassifies negative samples near the baseline as non-informative, leading to an overemphasis on the importance of positive samples.

\textbf{$\cdot$ Approximation Bias}. Gradient-based methods introduce errors due to the first-order Taylor approximation, which assumes minimal distance between the target and baseline features. However, in real-world scenarios, these distances are often significant, causing substantial approximation errors. Figure~\ref{Approximation_bias} highlights these errors, which previous methods largely overlook.

%% file: sections/method.tex
\begin{figure*}[!t]
  \centering
    \includegraphics[width=0.98\linewidth]{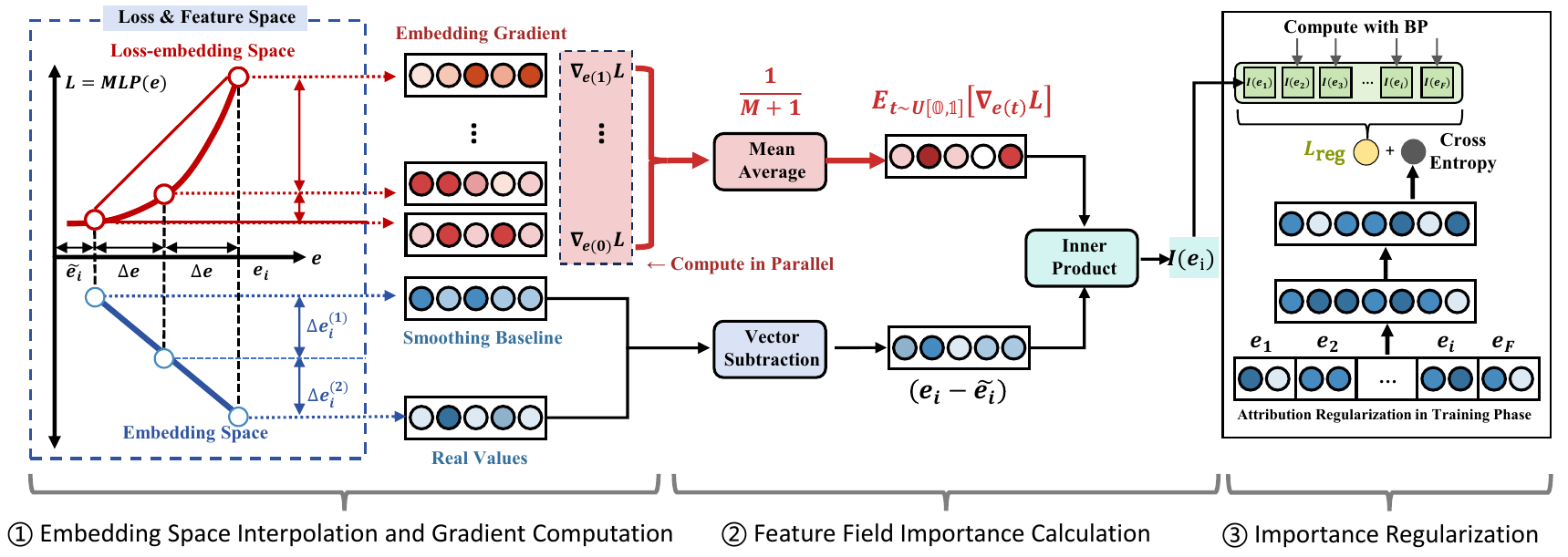}
  \caption{The FairFS framework consists of three stages: \textbf{Stage 1} sets anchor embedding points between the baseline and original feature embeddings, and computes their gradients. \textbf{Stage 2} combines these anchor points with their gradients to estimate final feature importance. ($M$ is in training phase while $n_{ac}$ is in validation phase) \textbf{Stage 3} illustrates how feature importance is applied in both the training and validation phases: in training, it serves as a regularizer, while in validation, it guides feature selection.}
  \label{fig:framework}
\end{figure*}

\section{Methodology}\label{sec:Model}
To mitigate the three bias issues present in previous methods, we propose FairFS. The overall framework of FairFS is shown in Figure~\ref{fig:framework}.
The framework includes a feature importance estimation module and an importance regularization module,
which can jointly reduce the number of feature fields the model relies on. Further details of our approach are provided below.
\subsection{Problem Definition}
The primary goal of feature selection in recommendation systems is to minimize the cardinality of the field subset while maximizing model precision. If we represent each feature field, such as \textit{gender} and \textit{age}, with $\boldsymbol{e_i}$ and aim to minimize the cross-entropy loss, the objective can be formulated as follows:
\begin{equation}
\begin{aligned}
    \underset{E_s}{\text{argmin}}\ &\ \text{card}(\boldsymbol{E_s} \,|\, \boldsymbol{E_s} \subset \boldsymbol{E}) \\
    & \text{s.t.}\ \ L(y, f_\theta(\boldsymbol{E_s})) - L(y, f_\theta(\boldsymbol{E})) < \delta,
\end{aligned}
\label{eql}
\end{equation}
where $\boldsymbol{E}$ and $\boldsymbol{E_s}$ represent full field set and field subset. 
$f_\theta$ represents the deep model parameterized by $\theta$. 
And $L$ denotes the cross-entropy loss between user's actual click behavior $y$ and the predictions made by the neural network.
Finally, this optimization problem seeks to identify the smallest possible subset of fields, $E_s$, that results in a decrease in model performance, compared to using the full set $E$, by no more than $\delta$, where $\delta$ is the maximum tolerable performance degradation as defined by AI engineers.

To quantify the contribution of each feature field, a feature importance metric needs to be established first. Based on this metric, the top-$K$ scoring feature fields will be selected within $E_s$ and utilized for model retraining. If the feature importance metric is denoted as a scalar function as $I(\boldsymbol{e_i})$ for the field $i$, then the objective function in Equation~\ref{eql} can be rewritten as follows:
\begin{equation}
\begin{aligned}
\text{argmin}_{E_s}\ \text{card}&\left(\boldsymbol{E_s} \mid \boldsymbol{E_s} = \{\boldsymbol{e_j} \mid \boldsymbol{e_j} \in \text{top-}K(I(\boldsymbol{e_i})), \right.\\ 
&\left.\forall \boldsymbol{e_i} \in \boldsymbol{E}\}\right).    
\end{aligned} 
\label{eq2}
\end{equation}

Then, the problem becomes how to accurately estimate the feature importance metric function $I$. 

\subsection{Gradient-based feature importance}
FairFS defines feature importance as the disparity in model performance when an informative feature $\boldsymbol{e_i}$ is replaced by an non-informative baseline feature $\tilde{\boldsymbol{e_i}}$ as below:
\begin{equation}
    I(\boldsymbol{e_i}) = 
    L(y, f_\theta(\boldsymbol{E}, \boldsymbol{E_i}=\boldsymbol{e_i})) - 
    L(y, f_\theta(\boldsymbol{E}, \boldsymbol{E_i}=\tilde{\boldsymbol{e_i}})).
    \label{eq3}
\end{equation}

This definition aligns with that 
of other sensitivity-based  algorithms mentioned in Section \ref{Related} like Permutation 
 \cite{permutation}, SFS \cite{multisfs}, and SHARK \cite{shark}, yet different in employing an unbiased baseline feature and more precise approximation methods.

Computing the Equation~\ref{eq3} requires $O(F)$ evaluation or training time complexities since it requires recalculating $L(y, f_\theta(\boldsymbol{E}, \boldsymbol{E_i}=\tilde{\boldsymbol{e_i}}))$ for each $i$.  
To accelerate this process, a straightforward way would be 
approximating $I(\boldsymbol{e_i})$ through a $1^{st}$ order Taylor expansion, multiplying the gradient at $\boldsymbol{e_i}$ by the vector distance between the original feature and baseline features, as illustrated below:
\begin{equation}
    I(\boldsymbol{e_i}) \approx 
    <\nabla_{\boldsymbol{e_i}} L(y, f_\theta(\boldsymbol{E}, \boldsymbol{E_i}=\tilde{\boldsymbol{e_i}})), 
    (\boldsymbol{e_i} - \tilde{\boldsymbol{e_i}})>.
    \label{eq4}
\end{equation}

This approximation method is highly efficient, which enables the parallel computation of all feature gradients and importance values using just one gradient operation and one dot product operation, thereby reducing the time complexity to $O(1)$. However, this approximation is rough and inaccurate due to the approximation bias of the Taylor expansion being $O(|\boldsymbol{e_i} - \tilde{\boldsymbol{e_i}}|)$, which escalates proportionally with the distance between $\boldsymbol{e_i}$ and $\tilde{\boldsymbol{e_i}}$. This bias underscores the necessity for the distance between $\boldsymbol{e_i}$ and $\tilde{\boldsymbol{e_i}}$ to be sufficiently small, ensuring that the approximation bias does not compromise the accuracy of feature importance estimation.
 
To mitigate the approximation bias, we implement two key strategies below. First, we \textbf{calculate gradient for loss} at different sampled points and \textbf{aggregate (average) them}, as detailed in Subsection \ref{Aggregated}. Second, we \textbf{design a baseline feature} that is closer to the original feature, which is discussed in Subsection~\ref{Smoothing}.

\subsection{Aggregated Feature Importance}
\label{Aggregated}
Drawing on the theoretical framework above, we determine the precise contribution of each feature to the loss degradation. While the exact computation requires a continuous evaluation, we approximate this value using a discrete sampling strategy inspired by the \textit{Mean Value Theorem for Integrals}. 
This theorem posits that the definite integral of a function over a path is equivalent to the product of the path length and the \textbf{average value} of the function along that path. Specifically, we define the continuous interpolation path $\boldsymbol{e}(t)$ from the input $\boldsymbol{e_i}$ to the baseline $\tilde{\boldsymbol{e_i}}$ parameterized by $t \in [0, 1]$ as:
\begin{equation}
    \boldsymbol{e}(t) = (1 - t) \boldsymbol{e_i} + t \tilde{\boldsymbol{e_i}},
    \label{eq:path}
\end{equation}
where $\boldsymbol{e}(0) = \boldsymbol{e_i}$ and $\boldsymbol{e}(1) = \tilde{\boldsymbol{e_i}}$.

According to the theorem, for this continuous path, there exists a specific mean point $c \in [0, 1]$ such that the gradient at $c$ equals the \textbf{average gradient} along the path. Mathematically, this average is formulated as the definite integral, which corresponds to the expectation under a uniform distribution:
\begin{equation}
    \int_{0}^{1} \nabla_{\boldsymbol{e}(t)} L \, dt = \nabla_{\boldsymbol{e}(c)} L = \mathbb{E}_{t \sim U[0, 1]} [\nabla_{\boldsymbol{e}(t)} L].
    \label{eq:mvt}
\end{equation}

Since the exact location of $c$ is unknown, we estimate the gradient at this point using the theoretical expectation term. Consequently, the feature importance is derived as:
\begin{equation}\small
\begin{aligned}
    I(\boldsymbol{e_i}) &= \underbrace{(\boldsymbol{e_i} - \tilde{\boldsymbol{e_i}})}_{\text{Feature Displacement}}\odot \quad \underbrace{\mathbb{E}_{t \sim U[0, 1]} [\nabla_{\boldsymbol{e}(t)} L]}_{\text{Theoretical Expectation}}.
\end{aligned}
\label{eq6}
\end{equation}

The theoretical expectation term represents the average gradient along the continuous path. To compute this value numerically, we employ a deterministic \textbf{equidistant sampling strategy} rather than stochastic Monte Carlo sampling. 
Specifically, we approximate the expectation using the arithmetic mean of gradients computed at $M+1$ anchor points as:
\begin{equation}
    \mathbb{E}_{t \sim U[0, 1]} [\nabla_{\boldsymbol{e}(t)} L] \approx \frac{1}{M+1} \sum_{m=0}^{M} \nabla_{\boldsymbol{e}(t_m)} L(y, f_\theta(\boldsymbol{e}(t_m))),
    \label{eq:expectation_approx}
\end{equation}
where\footnote{In the validation phase, we use $n_{ac}$ to represent the number of sampling steps replacing $M$.} $t_m = \frac{m}{M}$ and $\boldsymbol{e}(t_m)$ is the discrete anchor point defined in Equation~\ref{eq:path}.
We implement this strategy on the validation dataset as a post-training process. This decouples feature field selection from model training, effectively preventing overfitting to the training distribution. Given that our method provides \textbf{sample-wise} importance while feature selection requires a dataset-level metric, we calculate the global importance by summing the scores over the validation dataset to represent the feature field importance. As the point number increases, the approximation bias approaches zero, which is proven in Section~\ref{appro_proof}.

A limitation of the post-hoc method described above is that the model is not explicitly trained to optimize for feature sparsity. To integrate sparsity directly into the training process, we introduce the aggregated feature importance as a regularizer in the loss function.
As a simplification during training, for the regularization term specifically, we set $M=0$. This reduces the calculation to a first-order Taylor approximation at the input point (Gradient $\times$ Input), requiring virtually no extra computation beyond the standard training backward pass.
This proposed regularizer offers two key advantages over prior works:
1) \textbf{Non-linear Sensitivity:} Unlike AutoField~\cite{autofield} and MvFS~\cite{mvfs}, our regularization is based on gradients from the deep network, allowing it to capture feature importance through complex non-linear transformations and mitigate layer bias.
2) \textbf{Field-level Sparsity:} Unlike I-Razor~\cite{irazor} which operates on individual embedding dimensions, our approach aggregates importance at the feature field level, promoting sparsity for entire features collectively:
\begin{equation}
    L_{\text{reg}}(\boldsymbol{E}) = \lambda \sum_{i=1}^{F} ||I(\boldsymbol{e_i})||_2.
    \label{eq8}
\end{equation}
The full pseudocode for calculating aggregated feature importance is detailed in Appendix Algorithm~\ref{alg2}.

\subsection{Smoothing Baseline Feature}
\label{Smoothing}
The baseline feature $\tilde{\boldsymbol{e_i}}$ plays a pivotal role in gradient-based feature importance methods, serving as a fundamental assumption by identifying which type of feature is non-informative. Previous approaches such as Permutation, SHARK, and SFS have proposed fixed baseline features based on intuitive concepts like swap baseline, mean baseline, and zero baseline. However, these methods do not adequately articulate the Principle a baseline feature must meet or how their chosen baselines adhere to these Principles.

In our research, we delineate two essential principles that a baseline feature should satisfy:
(1) \textbf{Non-informative Principle}: This principle dictates that the feature should not possess information favoring any particular class in a binary classification task and should ideally be positioned exactly on the decision boundary between two classes.
(2) \textbf{Proximal Principle}: Given that the decision boundary is conceptualized as a hyper-manifold, any sample on this manifold satisfies the non-informative principle. To minimize the error from the first Taylor expansion, the baseline feature should be as proximal to the target sample as feasible.

Consequently, the problem of selecting an appropriate baseline can be expressed as the following optimization challenge:
\begin{equation}
    \min{\ (\boldsymbol{E} - \tilde{\boldsymbol{E}})^2,\ } s.t.\ f_{\theta}(\tilde{\boldsymbol{E}}) = D_b,
    \label{eq9}
\end{equation}
where the constraint term delineates the decision boundary manifold. In a typical binary classification problem with balanced positive and negative samples, this value should be 0 (before applying the Sigmoid function). However, in recommendation tasks, where negative samples outnumber positive ones, the decision boundary tends to be biased towards negative samples. This bias presents challenges in accurately measuring the baseline feature.

To compute the baseline feature, we made two simplifications. One is a two layer ReLU network assumption where $f_\theta(\boldsymbol{\tilde{\boldsymbol{E}}})=\boldsymbol{\Theta_1} ReLU(\boldsymbol{\Theta_2} \boldsymbol{\tilde{\boldsymbol{E}}}) + \boldsymbol{b}$ and the second is a balanced dataset assumption where $D_b=0$. Consequently, the above optimization problem can be reformulated as shown below with a Lagrange multiplier $\alpha$:
\begin{equation}
    min\ (\boldsymbol{E} - \tilde{\boldsymbol{E}})^2 + \alpha f_{\theta}(\tilde{\boldsymbol{E}})^2
    \label{eq10}.
\end{equation}

When $\boldsymbol{\Theta_2} \boldsymbol{\tilde{\boldsymbol{E}}}$ is greater than zero and ReLU is activated, by setting the first derivative of the objective function w.r.t. $\boldsymbol{\tilde{\boldsymbol{E}}}$ as zero, we get:
\begin{equation}
\boldsymbol{\tilde{\boldsymbol{E}}} = \boldsymbol{{\boldsymbol{E}}} - \frac{\alpha}{2}(\boldsymbol{\Theta_1} \boldsymbol{\Theta_2})^T,
    \label{eq11}
\end{equation}
which tells us each sample should have its unique baseline feature and the sample’s original information should be smoothed out by the inverse direction of neural network parameter. 
By iterating this gradient descent formula until $f_\theta(\tilde{\boldsymbol{E}}) = 0$, the most proximal baseline feature $\tilde{\boldsymbol{E}}$ can be grabbed from sample $\boldsymbol{E}$.

For simplicity, to efficiently surrogate the iterative optimization, we introduce a sample-wise \textit{Smoothing} operator. We interpret the input embedding $\boldsymbol{E}$ as a concatenation of $F$ distinct field representations, denoted as $\{\boldsymbol{h}_k\}_{k=1}^F$. 
The smoothing baseline is derived by computing the expectation of these sub-representations over \textbf{different} fields \textbf{within the same sample}, denoted as $\mathbb{E}_{k}[\boldsymbol{h}_k]$. This expectation captures the sample's intrinsic intensity while neutralizing field-specific semantics. The final baseline $\tilde{\boldsymbol{E}}$ is constructed by concatenating $F$ replicas of this expectation vector:
\begin{equation}
    \tilde{\boldsymbol{E}} = \Big[\ \underbrace{\mathbb{E}_{k}[\boldsymbol{h}_k],\ \mathbb{E}_{k}[\boldsymbol{h}_k],\ \dots,\ \mathbb{E}_{k}[\boldsymbol{h}_k]}_{F \text{ replicas}}\ \Big].
    \label{eq12}
\end{equation}
By universally replacing individual field representations with unique statistical properties of each sample, we effectively approximate the goal shown in Equation~\ref{eq10} and make the baseline feature (embedding) of each sample different.

\subsection{FairFS Algorithm}
FairFS directly specifies feature importance during training and delivers a final feature importance score in the evaluation phase to avoid overfitting on the training dataset. The final feature selection procedure is described in the Algorithm \ref{alg1}.

\begin{algorithm}[!t]
	\caption{FairFS} 
	\begin{algorithmic}
	    \STATE \textbf{Input}: Feature fields embedding $E$, surrogate model $f_\theta$, interpolation hyper-parameter $M=0$ and $n_{ac}$ ($M < n_{ac}$), loss combination hyper-parameter $\lambda$.
	    \STATE \textbf{Output}: Feature importance score
            \STATE \textbf{Training Phase:}
		\WHILE{Sample B in Training Dataset}  
		\STATE Conduct feed-forward computation $L(y, f_\theta(\boldsymbol{E}))$
		\STATE Compute feature importance as Algorithm~\ref{alg2} with $M$
            \STATE Compute importance regularization loss with Equation~\ref{eq8}
		\STATE Combined with cross-entropy loss and conduct optimization
		\ENDWHILE 
           \STATE \textbf{Validation Phase:}
           \WHILE{Sample B in the Validation Dataset}  
		\STATE Compute feature importance as Algorithm \ref{alg2} with $n_{ac}$
            \STATE Aggregate feature importance for each batch samples
		\ENDWHILE 
	\end{algorithmic} 
\label{alg1}
\end{algorithm}

%% file: sections/experiments.tex
\section{Experiments}
In this section, we will describe the performance assessment of FairFS. We use different feature selection methods from various perspectives to conduct this evaluation. Our study focuses on the following four questions:

\begin{itemize}[leftmargin=*]
\item \textbf{RQ1:} Can FairFS perform better than existing methods?
\item \textbf{RQ2:} Is FairFS efficient enough in practical usage?
\item \textbf{RQ3:} How do hyper-parameters affect FairFS?
\item \textbf{RQ4:} How does each module contribute to the final performance?
\item \textbf{RQ5:} Does FairFS's online performance align with the offline?

\end{itemize}

\subsection{Experimental Setup}

\begin{table*}[htbp]\small
  \centering
  \caption{Performance comparison between FairFS and other feature selection methods.}
  \vspace*{-0.3cm}
  
  \setlength{\tabcolsep}{3.5pt} 
  \begin{tabular}{cccccccccccccc}
    \hline\hline
    \multirow{2.5}{*}{Dataset}& Model & \multicolumn{6}{c}{Wide\&Deep} &
    \multicolumn{6}{c}{DCN}        \\\cmidrule{3-8}\cmidrule(l){9-14}
    & Method & No Selection  & PFI & AutoField & AdaFS & MvFS & FairFS & No Selection & PFI & AutoField & AdaFS & MvFS & FairFS \\
    \specialrule{0.1em}{0pt}{0pt}
    \multirow{3}{*}{Criteo} & AUC$\uparrow$ & 0.8090 & 0.8093 & 0.8093 &  0.8116 & \textbf{0.8120} & 0.8094 & 0.8095 & \textbf{0.8098} & 0.8096 & 0.8095 & 0.8097 & 0.8097\\
    & Logloss$\downarrow$ & 0.4425 & 0.4424 & 0.4423 & 0.4402 & \textbf{0.4398} & 0.4422 & 0.4423 & \textbf{0.4419} & 0.4432 & 0.4421 & 0.4420 & 0.4420\\
    & Ratio$\downarrow$ & 100.00\% & 97.44\% & 94.87\% & 97.44\% & \textbf{84.61\%} & 92.31\% & 100.00\% & \textbf{92.31\%} & 97.44\% & 97.44\% & \textbf{92.31\%} & \textbf{92.31\%} \\
    \specialrule{0.1em}{0pt}{0pt}
    \multirow{3}{*}{Avazu} & AUC$\uparrow$ & 0.7922 & 0.7923 & 0.7719 & 0.7877 & 0.7910 & \textbf{0.7929*} & 0.7920 & 0.7918 & 0.7920 & 0.7867 & 0.7915 & \textbf{0.7928*} \\
    & Logloss$\downarrow$ & 0.3726 & 0.3723 & 0.3840 & 0.3725 & 0.3735 & \textbf{0.3721} & 0.3726 & 0.3727 & 0.3732 & 0.3757 & 0.3732 & \textbf{0.3722*} \\
    & Ratio$\downarrow$ & 100.00\% & 95.83\% & \textbf{54.17\%} & 95.83\% & 58.33\% & 66.67\% & 100.00\% & 91.67\% & 95.83\% & 95.83\% & 70.83\% & \textbf{62.50\%} \\
    \specialrule{0.1em}{0pt}{0pt}
    \multirow{3}{*}{iFly-AD} & AUC$\uparrow$  & 0.8840 & 0.8854  & 0.8859 & 0.8857 & 0.8846 & \textbf{0.8862} & 0.8849 & 0.8852 & \textbf{0.8860} & 0.8859 & 0.8855 & 0.8852\\
    & Logloss$\downarrow$  & 0.04086 & 0.04077 & 0.04070 & 0.04074 & 0.04078 & \textbf{0.04067} & 0.04074 & 0.04071 & \textbf{0.04063}& 0.04071& 0.04071 & 0.04073\\
    & Ratio$\downarrow$  & 100.00\% & 81.63\% & \textbf{32.65\%} & 85.71\% & 93.88\% & 77.55\% & 100.00\% & 24.49\% &28.57\% & 77.55\% & \textbf{20.41\%} & 81.63\%\\
    \hline\hline
    \multirow{2.5}{*}{Dataset}& Model & \multicolumn{6}{c}{DeepFM}     &
    \multicolumn{6}{c}{FM}  \\\cmidrule{3-8}\cmidrule(l){9-14}
    & Method & No Selection                   & PFI & AutoField & AdaFS & MvFS & FairFS & No Selection & PFI & AutoField & AdaFS & MvFS & FairFS \\
    \specialrule{0.1em}{0pt}{0pt}
    \multirow{3}{*}{Criteo} & AUC$\uparrow$ & 0.8076 & 0.8072 & 0.8075 & 0.8071 & 0.8076 & \textbf{0.8086*} & 0.8010 & 0.8011 & 0.8006 & 0.8000 & 0.8007 & \textbf{0.8018*}\\
    & Logloss$\downarrow$ & 0.4444 & 0.4445 & 0.4445 & 0.4445 & 0.4441 & \textbf{0.4431*} & 0.4502 & 0.4502 & 0.4507 & 0.4510 & 0.4507 & \textbf{0.4497}\\
    & Ratio$\downarrow$ & 100.00\% & 97.44\% & 89.74\% & 97.44\% & 97.44\% & \textbf{92.31\%} & 100.00\% & 94.87\% & 92.31\% & \textbf{87.18\%} & 94.87\% & 92.31\%\\
    \specialrule{0.1em}{0pt}{0pt}
    \multirow{3}{*}{Avazu} & AUC$\uparrow$ & 0.7929 & 0.7931 & 0.7929 & 0.7917 & 0.7921 & \textbf{0.7932*} & 0.7828 & 0.7832 & 0.7830 & 0.7829 & 0.7830 & \textbf{0.7835}\\
    & Logloss$\downarrow$ & 0.3722 & 0.3719 & 0.3721 & 0.3729 & 0.3727 & \textbf{0.3718} & 0.3784 & 0.3781 & 0.3783 & 0.3782 & 0.3781 & \textbf{0.3779}\\
    & Ratio$\downarrow$ & 100.00\% & 95.83\% & 95.93\% & 95.83\% & 91.67\% & \textbf{83.33\%} & 100.00\% & 95.83\% & 95.83\% & 95.83\% & 91.67\% & \textbf{70.83\%} \\
    \specialrule{0.1em}{0pt}{0pt}
    \multirow{3}{*}{iFly-AD} & AUC$\uparrow$  & 0.8849 & \textbf{0.8855} & 0.8851 & 0.8846 & 0.8848 & \textbf{0.8855} & 0.8820 & 0.8845 & 0.8854 & 0.8855  & \textbf{0.8858} & 0.8843\\
    & Logloss$\downarrow$  & 0.04081 & 0.04074 & 0.04080 & 0.04084 & 0.04081 & \textbf{0.04073} & 0.04098 & 0.04082 & 0.04084 & 0.04096 & 0.04092 & \textbf{0.04078*}\\
    & Ratio$\downarrow$  & 100.00\% & 73.47\% & \textbf{20.40\%} & 32.65\% & 97.96\% & 61.22\% & 100.00\% & 77.55\% & 8.16\% & \textbf{4.08\%}  & 53.06\% & 69.39\% \\
    \hline\hline
  \end{tabular}
  {\caption*{\small \normalfont The symbol * denotes the significance level with $p \leq 0.05$. \textbf{Bold} font indicates the best-performing method.}}
  \label{tb:main_results}
  \vspace{-2.0em}
\end{table*}

In this section, we provide a detailed introduction to the implementation details. We conducted our experiments using the publicly available FuxiCTR repository\footnote{https://github.com/reczoo/FuxiCTR}, which offers recommended default parameters for all backbone models. To ensure a fair comparison, all feature selection methods were evaluated under the same backbone model parameters. Specifically:
\begin{itemize}[leftmargin=*]
\item 
\textbf{General hyper-parameters}: For three datasets, the training batch size was set to \textbf{10,000}. 
For Wide\&Deep \cite{wideanddeep}, DCN \cite{dcn}, DeepFM \cite{deepfm}, and FM \cite{fm} models, we adopt the default configuration from the repository, with embedding sizes of 40, 32, 40, and 10, respectively. To ensure that each model was adequately trained, we employ the same early stopping strategy and set the maximum training epochs to 100. Additionally, Adam optimizer~\cite{adam} and Xavier initialization~\cite{glorot2010understanding} are utilized.
\item 
\textbf{FairFS hyper-parameters}: Our method introduces two hyperparameters, \(\lambda\) and \(n_{ac}\), which control the sparsity of the gradient and the number of anchor points, respectively. For \(\lambda\), our search range is \(\{1e2, 1e1, 1, 1e-1, 1e-2, 0\}\); for \(n_{ac}\), the search range is \(\{1, 5, 10, 20\}\). The feature importance estimated is conducted on validation set to avoid overfitting.
\end{itemize}

\subsubsection{Comparative Methods}
We compared our FairFS method to baselines from following categories~\cite{jia2024erase}: Sensitivity based methods (PFI \cite{permutation}, SHARK\cite{shark}) and Gating-based methods (AutoField \cite{autofield}, AdaFS~\cite{adafs}, and MvFS \cite{mvfs}).

\subsubsection{Datasets}
We experimented with three public datasets, including Criteo, Avazu, and iFly-AD.
The data statistics are shown in Table~\ref{tab:data_stats}. We divide the training, validation, and test data sets in an 8:1:1 ratio. The model with the best performance in the validation set will be evaluated on the test set as performance results. 
Early stopping is also determined based on the performance on the validation set.
\begin{table}[htbp] 
  \caption{Detailed statistics of the three datasets.}
    \centering
  \label{tab:data_stats}
  \begin{tabular}{lcccc}
    \toprule
    \textbf{Dataset} & \textbf{\#Fields} & \textbf{\#Training} & \textbf{\#Validation} & \textbf{\#Test}\\
    \midrule
    Criteo & 39 & 36,672,493 & 4,584,062&4,584,062\\
    Avazu & 24 & 32,343,172 & 4,042,897 & 4,042,898 \\
    iFly-AD & 245 & 1,775,629 & 221,954 & 221,954 \\
  \bottomrule
\end{tabular}
\end{table}

\subsubsection{Base Models}
The backbone models we employ include: (i) \textbf{Wide\&Deep}~\cite{wideanddeep}: it combines linear models and deep neural networks to capture both direct relationships among sparse features and high-order combinations of complex features; (ii) \textbf{DCN}~\cite{dcn}: it integrates deep neural networks with explicit cross-network layers to capture high-order feature interactions and enhance predictive performance; (iii) \textbf{DeepFM}~\cite{deepfm}: it uses factorization machines and deep neural networks to explore interaction signals; (iv) \textbf{FM}~\cite{fm}: it adopts matrix factorization techniques to efficiently handle high-dimensional sparse data.

\subsubsection{Evaluation Metrics}\label{sec:Metrics}
Following the previous work~\cite{deepfm,fm}, we assess the results of feature selection using AUC, Logloss, and the ratio. Specifically, AUC is the area under the ROC curve, Logloss is the cross-entropy loss between projected outcomes and true labels, and the ratio is the percentage of selected fields number to total feature fields. A higher AUC value indicates a more effective selection procedure, while lower Logloss and ratio values indicate improved selection efficacy. It should be noted that in the task of CTR prediction, an improvement of \textbf{0.001} on AUC or logloss is considered significant \cite{deepfm,qu2016product}.
\begin{figure}[t!]
        \includegraphics[width=1\linewidth]{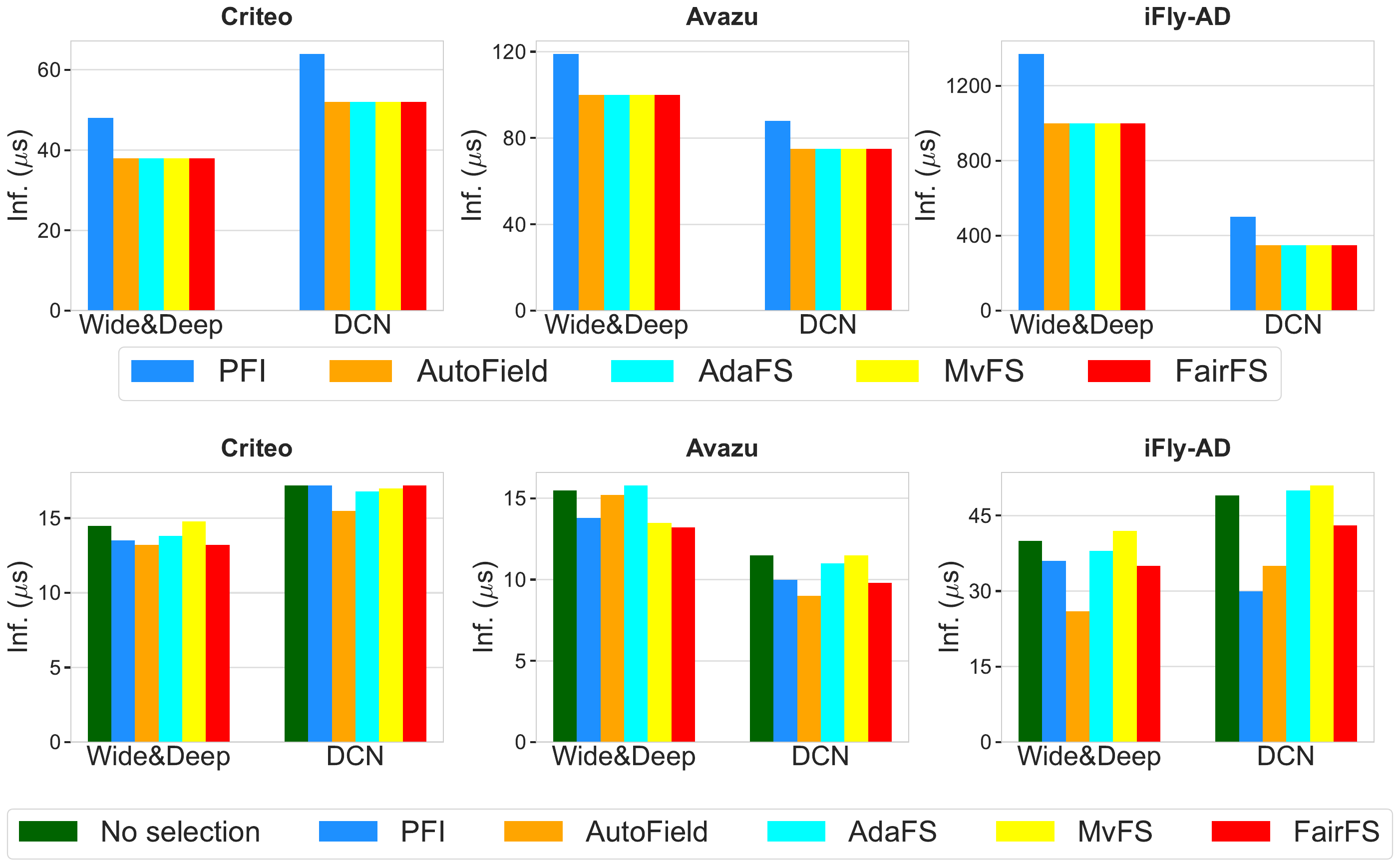}
    \caption{Efficiency comparison: feature selection time (Top) and inference time (Bottom) per sample}
    \label{fig:selection_time}
    
\end{figure}

\subsection{RQ1: Overall Performance}
We provide a detailed comparison between FairFS and other feature selection methods, as shown in Table~\ref{tb:main_results}. From the table, we can draw the following conclusions:

First, FairFS consistently improves the CTR model accuracy by eliminating redundant fields. On three datasets, FairFS significantly outperforms other methods, achieving relative improvements (RI) of up to 0.20\%, 0.22\%, and 0.35\% across the three datasets. Notably, on the iFly-AD dataset, although FairFS uses more feature fields than other baseline methods, it outperforms other methods on both Wide\&Deep and DeepFM models, demonstrating its ability to identify truly relevant features.
Second, FairFS effectively removes noisy fields, reducing training costs without sacrificing accuracy. For example, it eliminates about 30\% of redundant fields on the Avazu and iFly-AD datasets. In contrast, methods like MvFS, while removing more fields, also discard some informative ones, leading to accuracy drops. On the Avazu dataset, other methods achieve similar accuracy but require more fields, highlighting their inability to distinguish between redundant and informative features.
Finally, FairFS demonstrates stable selection performance. On the Criteo dataset, its accuracy remains optimal even when the field ratio reaches 90\%, and on both the Avazu and iFly-AD datasets, its performance fluctuations across backbone models are smaller than those of other methods. This stability indicates that, regardless of the backbone model, FairFS can overcome bias and consistently identify effective fields.

\subsection{RQ2: Efficiency Analysis}
For recommender systems, feature selection is an offline task triggered periodically when new features accumulate. Thus, an ideal algorithm should minimize serving latency (by selecting smaller feature subsets) while can tolerating longer offline selection times. We evaluate efficiency from both selection and serving perspectives.

\subsubsection{Selection efficiency}
Figure~\ref{fig:selection_time} shows the selection time per sample for different algorithms. FairFS is significantly faster than iterative methods like RFE and PFI but slightly slower than gating-based approaches. This minor overhead arises from the anchor embedding interpolation during validation. However, given that selected features are used long-term in online serving, this additional time is a reasonable trade-off for improved accuracy.

\subsubsection{Inference efficiency}
Inference efficiency is directly related to the size of the selected feature subset, reflecting the latency of using these features in online services. Figure~\ref{fig:selection_time} shows that FairFS achieves great inference efficiency across all datasets compared to no selection. In contrast, AdaFS and MvFS introduce additional controllers to assess feature importance for each record, which may increase inference costs. For instance, when using DCN on the iFly-AD dataset, these methods extend inference time. In summary, FairFS effectively selects a smaller but highly relevant feature subset, resulting in faster and more efficient inference while maintaining high prediction accuracy.

\begin{figure*}[!t]
    \centering
    \includegraphics[width=1\linewidth]{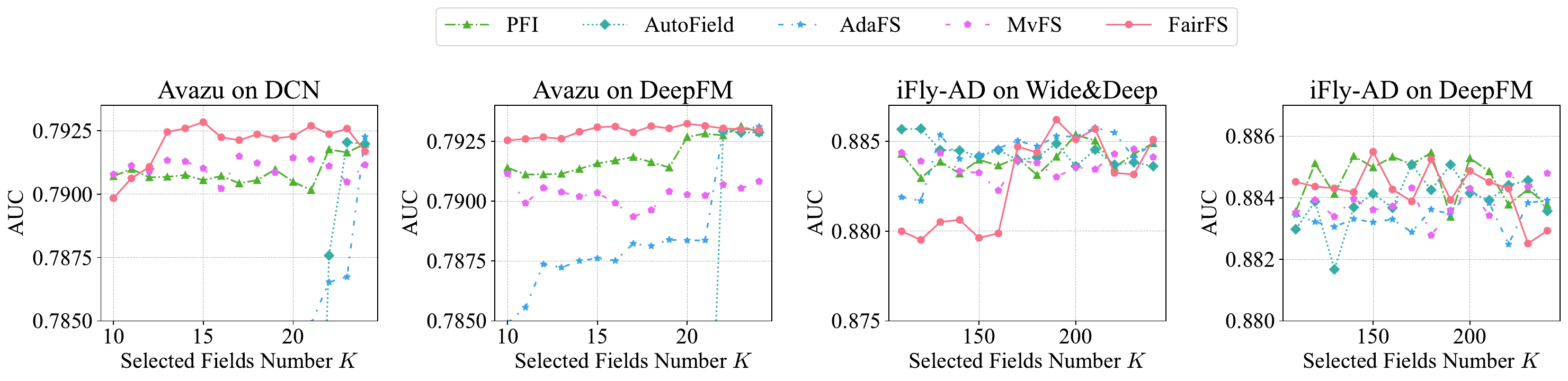}
    \caption{Parameter analysis for selected fields number $K$.}
    \label{fig:k_lines}
\end{figure*}

\subsection{RQ3: Hyperparameter Analysis}
\label{hyper}
\begin{figure}[!htb]
  \centering
  \begin{minipage}{0.23\textwidth}  
    \centering
    \includegraphics[width=\linewidth]{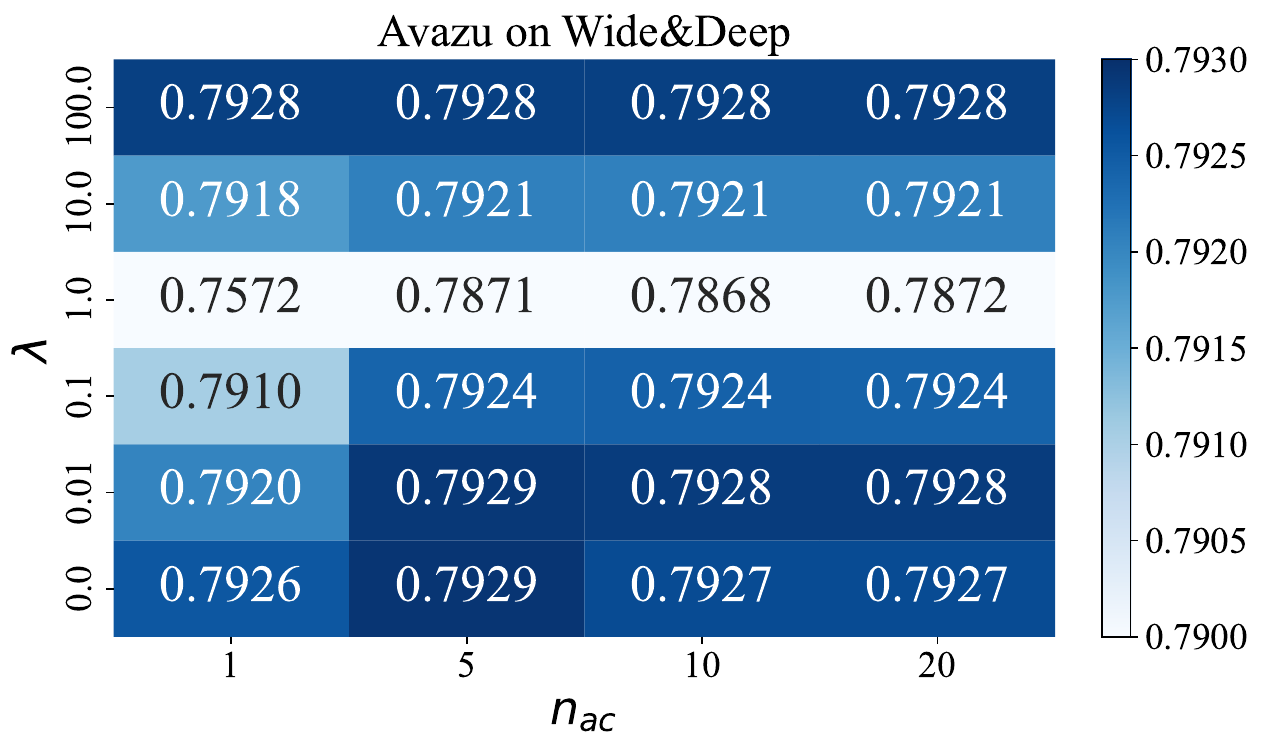}  
    \label{fig:image1}
  \end{minipage}\hfill  
  \begin{minipage}{0.23\textwidth} 
    \centering
    \includegraphics[width=\linewidth]{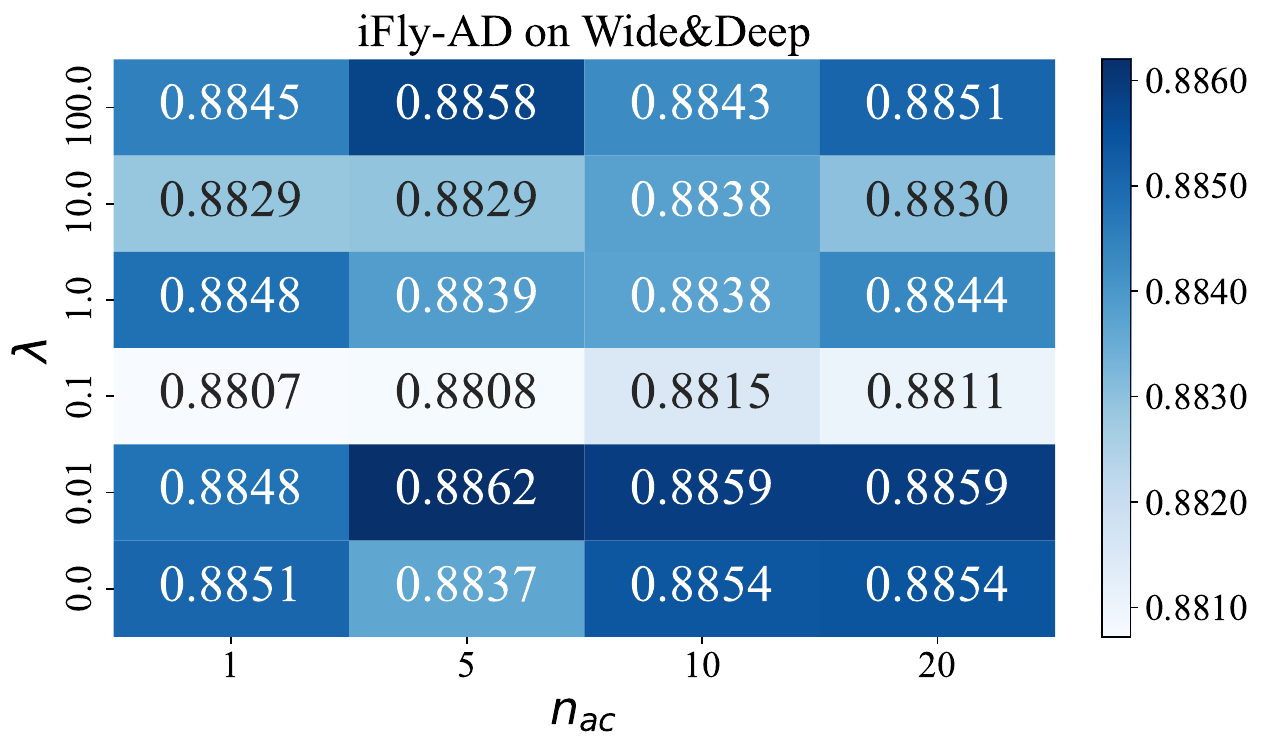} 
    \label{fig:image2}
  \end{minipage}
  \vspace*{-0.5cm}
  \caption{AUC under different hyper-parameter settings.}
  \label{fig:heat_map}
\end{figure}
We analyze three crucial hyper-parameters that influence model performance: the selected fields number \(K\), the number of anchor points \(n_{ac}\), and \(\lambda\), which controls the sparsity of the gradient

Firstly, Figure~\ref{fig:k_lines} shows how varying \(K\) affects AUC for Avazu and iFly-AD using DCN and DeepFM. On the Avazu dataset, FairFS consistently outperforms other methods when \(K > 13\). For DCN, the best performance is achieved when \(K < 20\), and for DeepFM, \(K = 20\) provides optimal results. This demonstrates that FairFS can effectively select useful features from a smaller subset of fields. Other methods, like AdaFS, require a larger \(K\) to perform well, while PFI does not perform as well as FairFS at the same \(K\), indicating it struggles to identify the most useful features. On the iFly-AD dataset, FairFS also achieves good selection performance.

Secondly, we analyze how \(n_{ac}\) and \(\lambda\) influence feature selection by varying them during training, while fixing \(K\) based on the values in Table~\ref{tb:main_results}. The AUC results are shown in Figure \ref{fig:heat_map}. As \(n_{ac}\) increases, the selection performance of FairFS generally improves, likely due to denser anchor points enabling finer and more accurate differentiation of feature fields. Meanwhile, \(\lambda\) performs best at lower values, indicating that excessively large regularization terms can lead to overlooking important feature fields.

\textbf{In industrial systems, practitioners can choose the maximum values for these parameters within computational limits.} However, \(\lambda\) should be carefully adjusted to balance sparsity and accuracy, with values between 0 and 0.01 providing the best results based on our study.

\subsection{RQ4: Ablation Analysis}
\label{Baseline_Analysis}
\subsubsection{Ablation Study on Aggregated Importance}
The ablation study on the aggregated feature importance module is conducted by setting the number of anchor points \(n_{ac}\) and regularization coefficient \(\lambda\) to zero, which reduces the feature importance calculation to a simple gradient-based selection with a smoothing baseline. As shown in Figure~\ref{fig:heat_map}, increasing \(n_{ac}\) from 0 to 40 results in a nearly monotonic improvement in feature importance estimation accuracy.

\subsubsection{Ablation Study on Smooth Baseline} 
We conducted two experiments to assess the effectiveness of the smooth baseline. In the first, we removed the aggregated importance module by setting \(n_{ac}\) and \(\lambda\) to zero and compared the smooth baseline’s proximity to the decision boundary with other baseline methods. The second experiment evaluates how different baseline features affect feature selection accuracy.
\textbf{Baseline features} we compare include: 
\begin{itemize}[leftmargin=*]
    \item \textbf{Zero}. Feature embedding with all zero entries. (SNIP \cite{snip} ).
    \item \textbf{Mean}. The mean of all samples' feature embeddings. (SHARK~\cite{shark}).
    \item \textbf{Swap}. Randomly swaps feature embeddings across samples to break original feature relationships (PFI \cite{permutation}).
    \item \textbf{Uniform Baseline}. Randomly samples noisy feature embeddings from a uniform distribution between the maximum and minimum values of the embedding. (MASK~\cite{mask}).
\end{itemize}

\subsubsection{Baseline bias analysis}
\begin{figure}[t!]
  \centering
  \begin{minipage}{0.23\textwidth}  
    \centering
    \includegraphics[width=\linewidth]{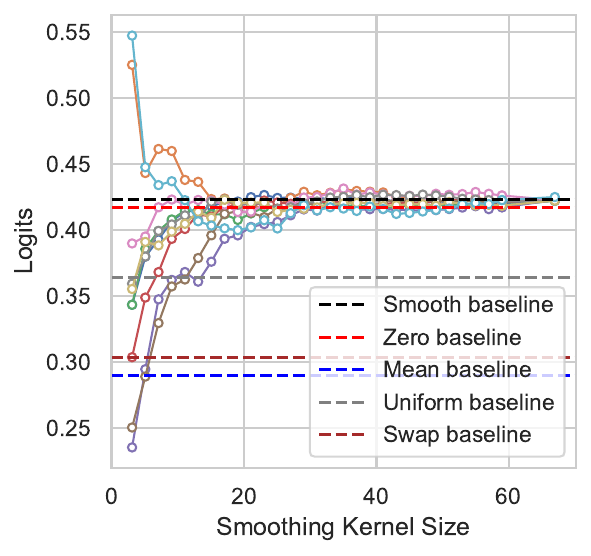} 
    \label{fig:image1}
  \end{minipage}\hfill  
  \begin{minipage}{0.23\textwidth}  
    \centering
    \includegraphics[width=\linewidth]{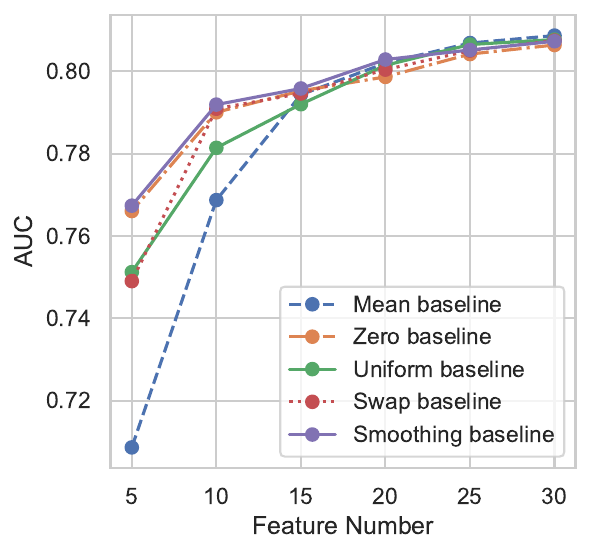} 
    \label{fig:image2}
  \end{minipage}
  \vspace*{-0.6cm}
  \caption{(Left) Case study on the baseline feature's informativeness. (Right) Model performance with different baselines. More biased baseline features lead to worse performance.}
  \label{fig:baseline_curve}
\end{figure}

We analyzed the feature selection performance with different baseline settings by fixing \(n_{ac} = 1\) and \(\lambda = 0\), and compared the accuracy on Criteo. The results, shown in Figure~\ref{fig:baseline_curve}, indicate that when the feature subset is large, the smooth baseline performs similarly to other baseline methods. However, with smaller feature sets, the smooth baseline exhibits more robust performance. This is because when the feature set is large, the removal of features due to data bias can be mitigated by the remaining features providing co-linear information. In contrast, for small feature sets, each feature is crucial, and the smooth baseline proves essential. Additionally, we performed a case study on the bias of each baseline by feeding the baseline features into a trained DCN model on Criteo. The output logits, representing the decision boundary positioning, revealed that both the smooth and zero baselines are closer to the decision boundary, with the smooth baseline having a shorter theoretical distance to the target feature embedding. Other baselines were more likely to be biased towards the negative side.

\subsection{RQ5: Online A/B Test} \label{online}
We deployed the feature selection process as an automatic pipeline. Once the number of newly engineered feature accumulated to 30 or the inference time of the current feature set exceed the maximum serving latency, the feature selection service will be invoked. Candidate features will be combined with the essence feature set (indispensable user and item features) as the input of feature selection algorithms. Then these features will be ranked and selected by their estimated feature importance for online serving.

We conducted one week's A/B online test on our industry ads platform, which can create tens of millions of daily income with over a hundred million users. 
We conducted two experiments to verify the effectiveness of FairFS on CTR and CVR scenarios respectively. 
For each experiment, we utilized FairFS to eliminate 23\% and 40\% noisy features respectively. The elimination ratio is determined by the feature selection curve. The selected feature subset is used as the experimental group while the feature subset selected by AutoField is used as the control group.
For each group, we assign 5\% traffic.
After deploying for one week, we observed a 1.35\% increase in ECPM and a 20\% decrease in online latency in the CTR scenario. In the CVR scenario, ECPM remained unchanged, while online latency decreased by 5.4\%. These results prove FairFS's capability in eliminating noisy features.
These two experiment groups' feature sets served the main traffic before new useful features were added.
Due to the unbiased property of FairFS, it has already become the feature selection baseline method in our platform.


%% file: sections/conclusion.tex
\section{Conclusion}
Noisy and redundant feature fields can increase the inference latency of recommender systems and may even lead to lower prediction accuracy.
To more precisely eliminate noisy features from deep recommender systems, in this work, we identified three bias issues, layer bias, baseline bias, and approximation bias, of current feature selection methods with demonstration of real-world examples. 
To avoid or mitigate these issues, we propose to directly regularize the aggregated feature importance during the training process with a smooth feature baseline.
We also give theoretical and experimental verification on how these methods can mitigate three bias issues.
Finally, extensive offline and online experiments show that the proposed FairFS method achieves robust and state-of-the-art performance in feature importance estimation, which highlight its potential practical value and applicability in real-world recommendation scenarios.


%% file: sections/appendix.tex
\appendix
\section*{Appendices}
\section{Notations}
\subsection{Feature Field}
\label{note}
A \textbf{feature field} refers to a specific column in the data table, representing the collection of values across all samples for a concrete attribute (\textit{e.g.}, \textit{City} or \textit{Device Type}). In this paper, our objective is to select a subset of informative fields to achieve superior prediction performance.

\subsection{Anchor Point Number}
The \textbf{Anchor Point Number}, denoted as $M$ (and $n_{ac}$), represents the count of discrete interpolation points used to approximate the gradient expectation. To balance computational efficiency with estimation precision, we adopt a phase-dependent strategy for this parameter: during the \textbf{training phase}, we set \textbf{$M=0$} to minimize overhead, effectively reducing the operation to a lightweight first-order Taylor expansion regularizer; conversely, during the \textbf{validation phase}, we use \textbf{$n_{ac}$} to represent it. As a strictly positive hyper-parameter (\textit{e.g.}, $n_{ac}=5$), it can help achieve a fine-grained and accurate assessment of feature field importance.

\section{Algorithm Pseudocode}\label{suppl:algorithm_code}
The algorithm for the feature field importance calculation.
\begin{algorithm}[ht!]
    \caption{Aggregated feature field importance}
    \label{alg2}
    \begin{algorithmic}
        \STATE \textbf{Input}: Data batch $(\boldsymbol{e}, y)$, Baseline feature $\tilde{\boldsymbol{e}}$ for each sample, Model $f_\theta$, Sampling steps $M$.
        \STATE \textbf{Output}: Aggregated feature importance score
        
        \STATE // 1. Define discrete anchor points
        \STATE Set $t_m = \frac{m}{M}$ for $m \in \{0, 1, \dots, M\}$
        
        \STATE // 2. Sample points along the path as in Equation~\ref{eq:path}
        \STATE Generate interpolated samples: $\boldsymbol{e}(t_m) = (1 - t_m) \boldsymbol{e} + t_m \tilde{\boldsymbol{e}}$
        \STATE Construct batch set $\mathcal{E} = \{ (\boldsymbol{e}(t_0), y), \dots, (\boldsymbol{e}(t_M), y) \}$
        
        \STATE // 3. Compute gradients via back propagation
        \STATE Conduct feed forward computation $L(y, f_\theta(\boldsymbol{e}(t_m)))$ for all samples in $\mathcal{E}$
        \STATE Compute gradients $\nabla_{\boldsymbol{e}(t_m)} L$
        
        \STATE // 4. Approximate expectation and compute importance as in Equation~\ref{eq6}
        \STATE Compute average gradient $\bar{g} = \frac{1}{M+1} \sum_{m=0}^{M} \nabla_{\boldsymbol{e}(t_m)} L$
        \STATE Calculate sample-wise importance $I(\boldsymbol{e}) = (\boldsymbol{e} - \tilde{\boldsymbol{e}}) \odot \bar{g}$
        
        \STATE // 5. Aggregate over the dataset
        \STATE Sum $I(\boldsymbol{e})$ over the validation batch to get field importance
    \end{algorithmic}
\end{algorithm}

\section{Approximation Bias Analysis} \label{appro_proof}

In this section, we analyze the convergence rate of our estimator. Recall from Equation~\ref{eq:mvt} that our theoretical target is the expected gradient height, which, by the \textit{Mean Value Theorem}, corresponds to the gradient at a specific point $c$:
\begin{equation}
    \mu = \mathbb{E}_{t \sim U[0, 1]}[\nabla_{\boldsymbol{e}(t)} L] = \nabla_{\boldsymbol{e}(c)} L.
\end{equation}

Here, we compare the estimation error $|\hat{\mu} - \mu|$ for two sampling strategies, where both methods utilize the same computational budget of $N_{samples} = M+1$ evaluations.
Consider a stochastic strategy where $M+1$ anchor points $t_0, \dots, t_M$ are sampled independently from a uniform distribution $U[0, 1]$. The estimator is the sample mean:
\begin{equation}
    \hat{\mu}_{MC} = \frac{1}{M+1} \sum_{m=0}^{M} \nabla_{\boldsymbol{e}(t_m)} L.
\end{equation}
Since the samples are i.i.d. random variables, according to the \textbf{Central Limit Theorem (CLT)}, the distribution of the estimation error converges to a Normal distribution:
\begin{equation}
    \sqrt{M+1} (\hat{\mu}_{MC} - \mu) \xrightarrow{d} \mathcal{N}(0, \sigma^2),
\end{equation}
where $\sigma^2$ is the variance of the gradient along the path. This implies that the expected error decays at the rate of the inverse square root of the sample size:
\begin{equation}
    \text{Error}_{MC} \propto \frac{1}{\sqrt{M+1}} \approx O(M^{-1/2}).
\end{equation}
This indicates that the estimation error approaches zero as $M \to \infty$.